# Adaptive Network Coding Schemes for Satellite Communications

Ala Eddine Gharsellaoui*, Samah A. M. Ghanem†, Daniele Tarchi* and Alessandro Vanelli-Coralli*
*Department of Electrical and Electronic Engineering
University of Bologna, Italy
†Huawei R&D Center, Sweden

***Abstract*—In this paper, we propose two novel physical layer aware adaptive network coding and coded modulation schemes for time variant channels. The proposed schemes have been applied to different satellite communications scenarios with different Round Trip Times (RTT). Compared to adaptive network coding, and classical non-adaptive network coding schemes for time variant channels, as benchmarks, the proposed schemes demonstrate that adaptation of packet transmission based on the channel variation and corresponding erasures allows for significant gains in terms of throughput, delay and energy efficiency. We shed light on the trade-off between energy efficiency and delay-throughput gains, demonstrating that conservative adaptive approaches that favors less transmission under high erasures, might cause higher delay and less throughput gains in comparison to non-conservative approaches that favors more transmission to account for high erasures.**

***Index Terms*—Adaptive Modulation Schemes; Energy Efficiency; Network Coding; Satellite Communications.**

## I. Introduction

5G wireless systems aim to 1000 fold capacity gains with 1 ms maximal latency, high quality of experience, seamless connectivity for users and global ubiquity. Such goals put forth challenges on introducing key enabling technologies to be integrated with wireless communications systems.

Since the work by Ahlswede et al. [1], which introduces network coding as a promising technology for wired networks, given the noiseless assumption of wired networks, adaptation techniques become less relevant. Thereafter, there have been several contributions that aim to study the performance gains of using network coding in different applications including wireless networks, associated with challenging characteristics that are relevant to proposals of adaptive schemes.

Network coding has been defined as a 5G key enabling technology for 5G wireless and wired networks. Several efforts are going towards standardizing such technology to current protocols and systems in practice [2]. The study of network coding for wireless networks need to take into consideration different aspects in the wireless medium like noise, interference, and fading.

The diversity benefits of network coding that can mitigate the wireless fading was shown in [3]. Moreover, from a decodability perspective, the authors in [4] proposed ZigZag decoding that is based on interference cancellation, and hence, requires a precise estimation of channel coefficients for each packet involved in a collision. In [5] an opportunistic network coding approach was introduced. In particular, the codewords are adapted according to the receiving information from the neighbors. In [6], the authors show that fixed network codes without channel state information (CSI) cannot achieve instantaneous min-cut. However, they proved that adaptive network codes with one bit global CSI have lower erasure probability than the codes without CSI.

Network codes adaptation strategies that are based on CSI are limited to packet erasure channel model, that is, a two-state Markov model of Gilbert-Elliott channel, in which a packet is whether dropped with a certain probability or received without error, see [7], [8]. In [7], the authors developed a rate-controlled, multipath strategy using network coding. They showed that such strategy can provide throughput performance comparable to multipath flooding of the network while utilizing bandwidth nearly as efficiently as single-path routing. In [8], the authors studied the delay and energy performance under bursty erasures. They proved that channel-aware policies reduce delay by up to a factor of 3 and significantly increase the network's stable throughput region compared to a simple queue-length driven policy. Further, the authors in [9] explored the timing nature of coding across packets over time division duplexing channels. However, they consider time invariant channels, i.e., with fixed erasure probability.

Of practical relevance is the study of channel variation and fading over time. Despite the fact that random linear network coding (RLNC) inherently adapts its rate to variations of the channel, however, in fading channels, the packet erasures become time dependent. The work by Ghanem in [10] stands as a milestone towards modeling coded and uncoded packet transmission over time varying channels and allows for proposals of network coding schemes that rely on the channel variation awareness and prediction to define strategies of packet transmission that allows for significant delay and throughput gains.

In this paper, we capitalize on the modeling framework in [10] to propose adaptive coding and modulation schemes for time varying channel. We compare such schemes with non-adaptive and adaptive network coded and uncoded schemes for time varying channels, applied to high speed satellite broadcasting scenarios.

Our goal is to exploit the channel variability on the fly to adapt the strategy of transmission of packets allowing for delay-throughput performance gains and energy efficiency which can mitigate the high Round Trip Time (RTT) of the satellite communications.

The contributions of this paper are: (i) we propose a novel adaptive network coding energy efficient (ANCEF) scheme for time varying channel that relies on less transmission under high erasures; (ii) we propose a modified adaptive network coding and modulation (ANCM) scheme that allows for limited number of retransmissions compared to uncoded schemes over equivalent number of transmission-retransmission time slots and adapts jointly its modulation scheme according the channel erasures; (iii) we utilize delay and throughput approximations to simulate Low Earth Orbit (LEO), Medium Earth Orbit (MEO) and Geostationary Earth Orbit (GEO) satellite scenarios characterized by different RTTs, establishing the fact that there is a clear trade-off between the packet size per transmission and energy efficiency or delay-throughput performance in the satellite scenario addressed; (iv) we established the fact that smart channel aware adaptive transmission strategies can be switched on and off alternatively to allow, through transmitter silence or larger batches of coded transmission, less delays, higher throughput and better utilization of the physical channel by inherent reduction of retransmissions.

The remainder of the paper is organized as follows: Section II introduces the channel model, a general review of packet transmission over time varying channel, and the delay or expected time to deliver the native or uncoded packets, Section III introduces the proposed adaptive network coding with energy efficiency scheme and adaptive coding and modulation scheme, and their corresponding delay approximation, while in Section IV the performance results are shown. Finally in Section V, the main conclusions are drawn.

## II. System Model

We consider a wireless communications system with time-variant channel. In such system, the transmitter performs RLNC, which is a NC scheme that relies on coding across the packets using random linear coefficients in order to increase the transmission reliability mimicking the wireless diversity concept.

Among several NC schemes proposed in the literature we resort to the system proposed in [10], where the author proposes a NC scheme based on the physical layer aware adaptation of batches of linearly combined packets whose number depends on the degree of freedom required at the receiver side.

Let's consider to have $N$ packets or degrees of freedom (dof) to be transmitted. The NC scheme can be exploited in such a transmission system by supposing the receiver able to send an ACK packet containing the lost degrees of freedom. This in turn can be exploited at the transmitter side for transmission of successive batch of packets whose length is equal to the remaining degrees of freedom yet required at the receiver side to allow for correct decoding of the native packets. Hence, if $N_i$ is the number of coded packets to be sent in batches for combining $i$ degrees of freedom, the transmission scheme will work for sending at each step $N_i \geq i$ packets, where $i$ is the number of degree of freedom to be sent, which is the number of independent linear combinations required to recover the transmitted packets via performing Gaussian elimination to solve the resulting linear system.

The process is repeated until all coded packets that are transmitted. Such scheme capitalizes on the proposed packet transmission Markov model for time varying channels in [10], where each state is represented by the couple $(i, h_j)$ that stands for the number of degrees of freedom to be sent and the channel state $h_j$, whose value can change during time.

Each two successive states are characterized by the transition probabilities between them, i.e., $p_{(i,h_j)\to(l,h_k)}$, that corresponds to the probability of correctly sending $i-l$ degrees of freedom, and the following state is $k-j$ time slots after the starting state. This means that $k-j$ corresponds to the number of channel variations along the transmission of the degrees of freedom which is $k-j \geq N_i$ and equals to $N_i+1$ if ACK is considered. The probability of each path outgoing from a certain state depends on several parameters, such as the modulation type, the channel gain, and the length of the packets.

We construct the one step transition probability matrix $P$, of a size defined by finite number of states considered in the Markov chain, and its components are defined with two transition probability components, $p_{(i,h_j)\to(i,h_{j+N_i})}$ and $p_{(i,h_j)\to(l,h_{j+N_i})}$ $\forall l < i$. The one step transition probabilities are:

$$p_{(i,h_j)\to(i-1,h_{j+1})} = 1 - P_e(h_j), \tag{1}$$

and

$$p_{(i,h_j)\to(i,h_{j+1})} = P_e(h_j), \tag{2}$$

where $P_e(h_j)$ is the packet erasure probability when the channel $h(t) = h_j$ for the duration of the packet transmission, and the probability of transitioning from the channel state and back to itself equals zero due to channel variation over time.

From a timing perspective, if the transmission of $N_i$ coded packets over $h_j, \ldots, h_{j+N_i}$ was not successfully decoded for all $N_i$ coded packets we account for the extra time via the probability of transitioning and being absorbed after first transmission, in addition to the time that accounts for the lost packets or lost dof at the receiver. Therefore, the expected time required to deliver $N_i$ coded packets is a sum of the first transmission time, the waiting time to have acknowledgment, and the transmission time of the lost dof; where the ACK packet includes the information of the lost dof, during which the channel process evolves until a new transmission takes place at the new channel state. Therefore, the expected time to deliver $N_i$ coded packets is given as follows,

$$T(i,h_j) = T_d(i,h_j) + \sum_{l=1}^{i} P^{N_i}_{(i,h_j)\to(l,h_{j+N_i})}\, T(l, h_{j+N_i+1}), \tag{3}$$

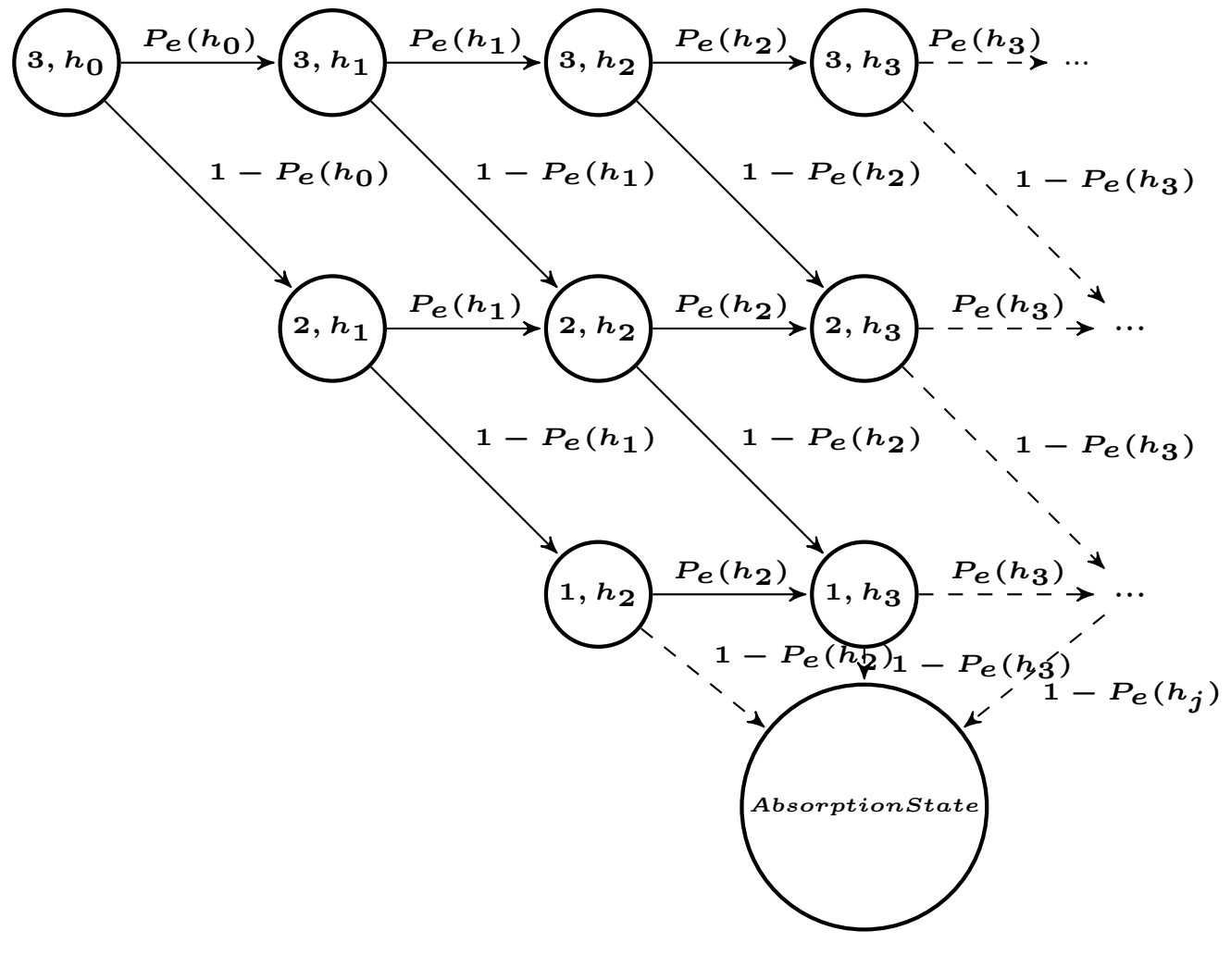

Fig. 1. Time Varying Channel Model of 3 Packets Transmission in [10]

with $T_d(N_i, h_j) = N_i T_p + T_w$, $T_w = Tp$ corresponds to the waiting time for acknowledgment. However,

$$\left(\prod_{i=1}^{N_i} P\right)_{(i,h_j)\to(l,h_{j+N_i})} = P^{N_i}_{(i,h_j)\to(l,h_{j+N_i})}, \quad (4)$$

corresponds to the $N_i^{th}$ step transition of matrix $P$. The $j + N_i + 1$ appears in the timing consideration due to acknowledgment.

The Markov chain as considered in [10] assumes a finite number of time slots to allow the transmission and reception of a given number of packets. Thus, the approximation of delay resorting to this model inherently constraints the number of re-transmissions of packets, but has sufficiently large number of slots for precise approximation.

By doing an example, represented in Fig. 1, where we suppose to transmit 3 packets, the first packet transmission at state $(3, h_0)$ will be either successfully delivered at channel state $h_0$ with probability $1 - P_e(h_0)$ causing a transition to occur to state $(2, h_1)$. However, the failure in its transmission corresponds to a probability $P_e(h_0)$ causing a transition to state $(3, h_1)$. In the following events the transition will occur either to state $(3, h_2)$ or to state $(2, h_2)$ in case of packet delivery failure or success, respectively. After a certain number of time slots, if the last packet at state $(1, h_k)$, $k \geq 3$ is delivered successfully, the process will be absorbed at a terminal state of the Markov chain, corresponding to end of transmission/reception process.

In the following subsection, the adaptive scheme proposed in [10] is described. This scheme paves the way towards our novel schemes, allowing for a fundamental question about the tradeoff between throughput and delay: shall we transmit more to obtain higher throughput? Or to transmit less to allow for energy efficency?

### A. Adaptive Network Coding (ANC)

The adaptive coded transmission scheme in [10] relies on the physical layer awareness of the packet erasures over the time varying wireless channel. Due to such variation in the packet erasure probability over each packet transmission, given by $P_e(h_j)$, the receiver can successfully decode $1 - P_e(h_j)$ packets. Therefore, the adaptive transmission strategy will be to account for the lost packets or lost dof, via the transmission of coded packets. The following equation presents the proposed strategy,

$$\sum_{s=j}^{N_i^*} (1 - P_e(h_s)) = i, \quad (5)$$

with $j = 0$ corresponds to the initial state the timing starts with for the $T(N_i, h_0)$. If $N_i$ packets transmitted starting from $h_0$ to receive successfully $i$ dof, then for each other transmission we account for the lost dof by extra coded packets. For example, if the receiver need to receive successfully $i$ dof, then the transmitter need to account for the packet erasures. This is done by successfully transmitting $N_i$ coded packets; at the point where the sum of $(1 - P_e(h_s))$ is equal to the dof, the process is repeated until the last dof is delivered to the receiver. The adaptive strategy will produce for each SNR a set of optimal number of coded packets to transmit, called $N_i^*$.

The implication of this process on the time is that the transition matrix should span up to $N_i^*$ in time slots. Therefore, we adaptively transmit accounting for packet erasures caused by channel variation over time as well as the dof of the receiver. Notice that the adaptation is controlled by the SNR level. Therefore, the time when to switch on or off adaptation is of particular importance to the process. In fact, adaptation is more required at the low-SNR regime. Furthermore, its worth to note that such adaptive scheme is of hybrid nature, through which it might sometimes behave as a selective repeat ARQ scheme when erasures are almost equivalent across all transmission/retransmission process or when the $N_i^*$ is constrained to a certain maximum value per transmission.

## III. Proposed Adaptive Network Coding Schemes

The concept of adaptive schemes is to transmit batches of packets dependently (as packet erasure become time dependent over time varying channels). In the two following schemes we are adapting and constructing batches of coded packets to allow for energy efficient schemes. Unlike the adaptive scheme in [10] whose main aim is to increase the throughput, in our proposed schemes, we either adapt the packets via less transmissions when erasures are high, or we utilize the one in [10] to transmit more to account for high foreseen erasure, while we yet add the energy efficiency concept by using lower modulation order when erasures are high, an enforced trade-off of particular relevance.

### A. Adaptive Network Coding with Energy Efficiency (ANCEF)

The goal of this scheme is to adapt the transmission to be energy efficient, by following the observation of the channel erasure; the strategy is to transmit small batches of coded packets if the observation of channel erasure is high (applies to low SNR), and to transmit more if erasure is less (applies to high SNR). The following equation illustrates number of coded packets $N_i$ required when $i$ dof are expected at the receiver,

$$\sum_{s=1}^{i}(1 - P_e(h_s)) = N_i \tag{6}$$

Therefore, this scheme will produce batches of low number of coded packets at greater values of channel erasure. Similarly, at low SNR with higher channel erasures, the number of retransmissions of the lost dof is also reduced. As the SNR increases associated with a vanishing channel erasures, the number of coded packets increases until it matches with the exact required $i$. In a similar way, the throughput at high SNR saturates at the level $1/T_p$.

### *B. Adaptive Network Coding and Modulation (ANCM)*

The second adaptive scheme we are considering is based on the exploitation of different modulation methods as foreseen in the most modern standards. By resorting to the adaptation rule exploited in the Adaptive Network Coding (ANC) scheme, our idea is to include adaptation of the modulation to the scheme. The rationale behind this, is that, on the one hand, a higher modulation order is more efficient, allowing for transmitting the same amount of information in shorter packets; on the other hand, due to the higher bit and symbol error probability, a higher number of packets could needed to be sent. This introduces a trade-off between the packet length and the number of coded packets for a given modulation scheme.

The proposed strategy is illustrated in the following equation:

$$\sum_{s=j}^{\min(N_i^*,\psi)} (1 - P_{e_m}(h_s)) = i, \tag{7}$$

where $P_{e_m}(h_s)$ is the erasure probability of the $m$-th modulation order that can be derived as:

$$P_{e_m} = 1 - (1 - P_{b_m})^B \tag{8}$$

where $P_b^m$ is the bit error probability for the same $m$-th modulation order while $B$ is the number of bits within one packet, that we consider as constant for fair comparison. $N_i^*$ corresponds instead to the number of coded packets to be sent in each transmission, which is optimized, and $\psi$ constraints the number of retransmissions to a value that does not increase the delay. The value of $\psi \leq N_i^*$ at low SNR, and $\psi > N_i^*$ at high SNR, at such range when erasures are vanishing, and $N_i = i$ almost surely.

Indeed the aim of the scheme is to find the optimal number of coded packets to transmit/re-transmit to assure successful reception of a given number of $i$ dof. Since a different modulation order corresponds to a different packet duration, the aim is to select the modulation order that allows for minimal length in time, of the batch of transmitted packets.

## IV. Numerical Results

We shall now provide a set of illustrates results that casts further insights on the proposed schemes. In particular, we focus on a satellite scenario by comparing the effectiveness of the schemes considering three orbital constellations, i.e., LEO, MEO, and GEO, resulting in different round trip times due to different transmitter to receiver distances.

To construct the simulation environment, we setup a maximum batch length in number of packets per transmission/retransmission to be constrained to a size $\psi$ equal to 10 packets. This is associated with a finite number of time slots per packet constrained to $\psi$ equal to 10 as well. This generally refers to a maximum number of total packets of $\psi^2$ equal to 100, which correspond to the size of the symmetric transition matrix.

The performance of the proposed schemes has been compared in terms of delay and gains, and with respect to its energy efficiency, comparing the number of packets per time unit each scheme can support.

To this aim the schemes have been compared by considering the transmission of 3 native packets. Our benchmark schemes are the ones proposed in [10] for time variant channels.

The first is the non-adaptive network coding scheme. In such scheme, it is clear that the number of coded packets are fixed along the transmission/retransmission with no adaptation considered. The second benchmark scheme is the adaptive network coding scheme. Contrary to the non-adaptive network coding scheme, the three adaptive schemes adapt the number of coded packets for each batch depending on the missing dof and on the channel erasures at a given estimated channel gains.

The modulation scheme considered is the BPSK in case of NC, ANC and ANCEF, while the ANCM exploits four possible modulation methods, i.e., BPSK, QPSK, 8PSK and 16QAM, in order to efficiently exploit the channel behavior. The modulation selection is based on the scheme previously described.

The number of bits per packet is considered 1000, while the maximum number of packets per batches to 10, leading a maximum batch of packet size that is equal to 10000 bits. Finally, as for the Satellite Channel, we consider a Ricean fading channel, that varies in time, with a $K$-factor equal to 10.

Before comparing the performance of the proposed adaptive network coding schemes in the three satellite scenarios, the average number of sent packets for the different schemes is compared, independently from the different RTT in the different scenarios; this can be seen in Fig. 2, where the average number of sent packets are represented for different average SNR values. As expected, it is possible to see that the ANCEF scheme has the lower number of sent packets, since it has been designed for limiting the number of sent packets in case of bad channel conditions for reducing the energy consumption.

In general, we can see that the average number of packets for all the schemes at low SNR is greater than those at high

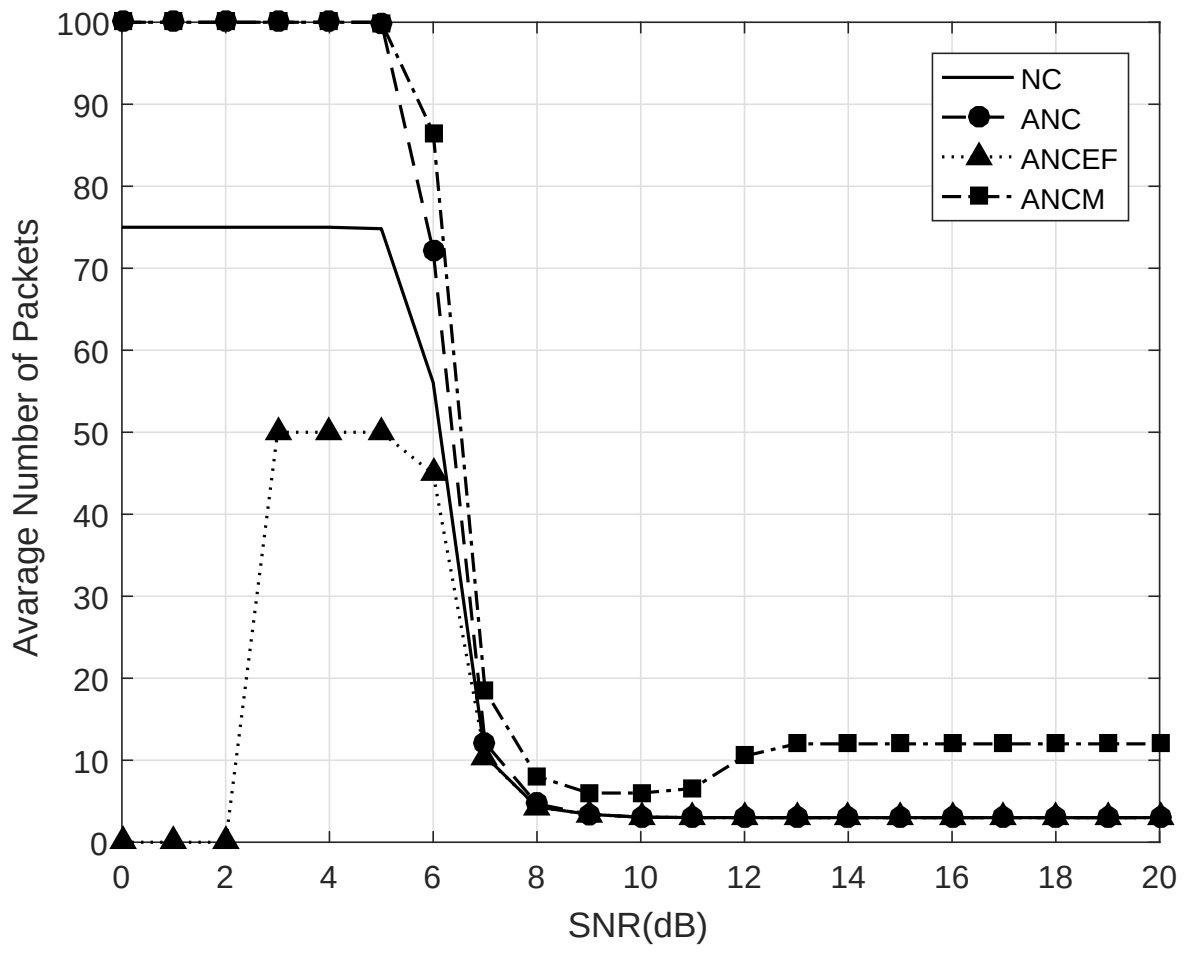

Fig. 2. Average number of sent packets for variable SNR values.

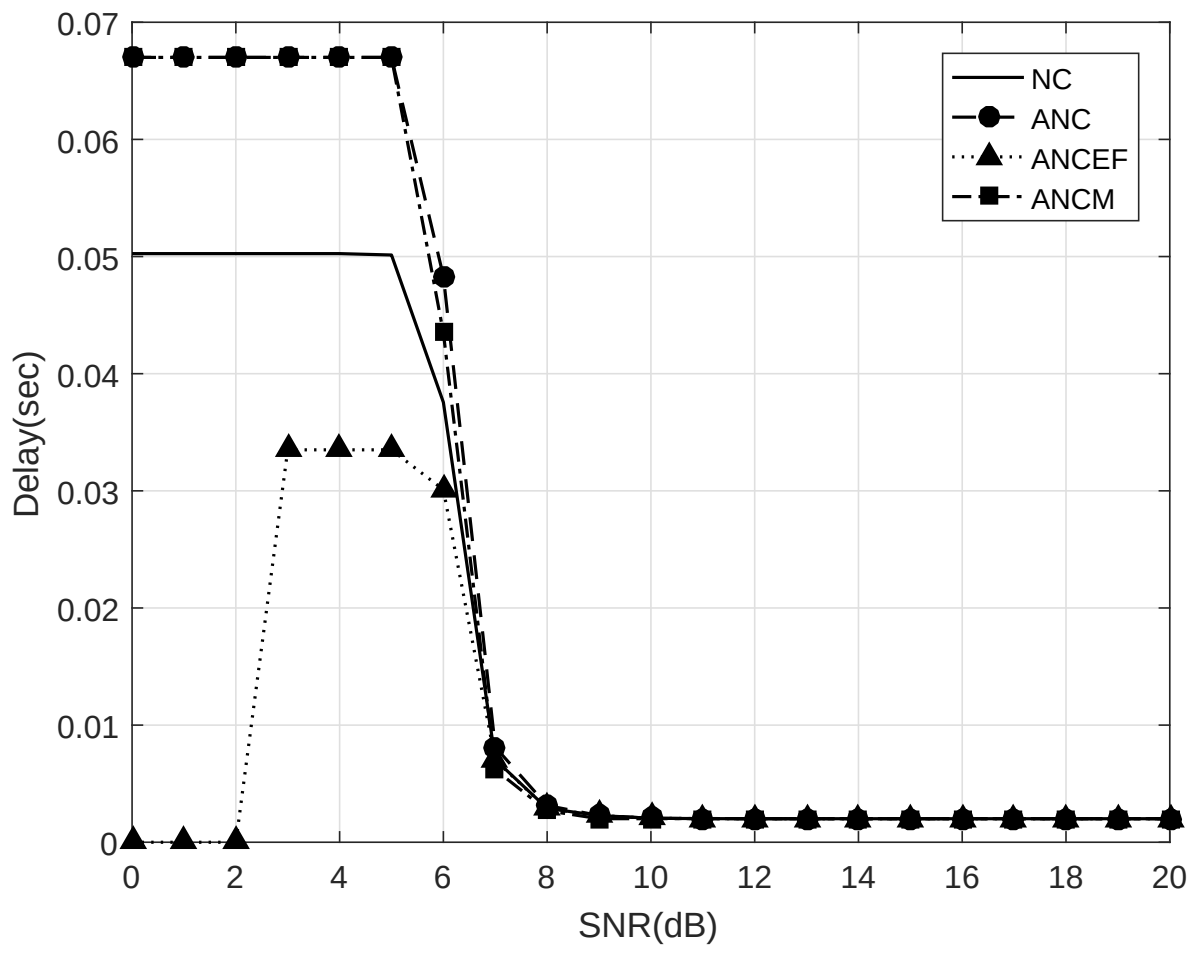

Fig. 3. Transmission Delay in an ideal scenario with RTT=0 for variable SNR values.

SNR due to the higher probability of re-transmission at low SNR. Digging into depth of each scheme and taking NC as a measure of comparison, since ANCM is a derived version of ANC where the selection of the modulation method is adapted, both schemes at low SNR values uses the maximum constrained size of a batch in order to reach the highest possible delivery, while in ANCEF due to energy efficiency, the average number of packets will be as much little as to commensurate with the higher erasure probability at low SNR i.e., there is no need to spend more energy in such low chance of delivery. This is even more highlighted at very low SNR, where the erasure probability is so high that the ANCEF foresees to not sending anything in order to avoid energy wasting.

It is worth to note that if both adaptive network coding schemes, ANCM and ANC are not constrained to a batch of maximum $\psi$ packets per transmission, then the delivery will be successful in fewer number of transmission/retransmission periods, defined by how far we can rely on the channel prediction before the CSI becomes obsolete. If the CSI is yet available, such adaptive schemes might achieve its highest possible delivery from a single shot transmission, with no need for repetition.

Moreover, it is worth to notice that for low SNR values ANCM employs the most robust modulation method (in terms of error probability), i.e., BPSK, resulting in the same performance of the ANC scheme.

Indeed, for higher SNR, corresponding to lower packet erasure probability, the average number of packets decreases then converges to the number of packets required at the receiver, or the dof value at the first transmission. It is possible to see that in ANC and ANCM the number of sent packets is higher than the NC scheme; this is expected since we aim to adapt the number of sent packets for achieving a target reliability. However, the maximum constrained batch size in ANCM is affected not only by SNR but also by modulation method, so at low SNR it behaves similarly to the ANC, then, going to higher SNR, it selects a higher modulation order allowing to increase the number of combinations into the level of code words or symbols that is transmitted over the channel, i.e., higher probability to achieve the same number of degree of freedom in that modulation order, cast to the exploitation of higher diversity for higher modulation order.

The effectiveness of the ANCEF scheme in terms of energy consumption can be also seen in Fig. 3, where the delay is represented by considering an ideal scenario where the RTT is equal to 0. Indeed it is possible to see that the delay for the ANCEF is the lowest with respect to the other schemes. A final remark on Fig. 3 that relates to the delay of the ANCM scheme which decreases as the SNR increases meeting with other schemes at the lowest saturated delay that is equal to the transmitted packet length in time. This in fact is due to the fact that the packet length is shorter for the higher modulation orders used at high SNR.

In Fig. 4, we can see the behavior in terms of transmission delay for the proposed schemes in a LEO satellite scenario.

It is possible to see that ANCEF has the higher delay for low SNR values; this is due to the fact that the scheme adapts its transmission to small size batches at the low SNR associated with high erasures. Therefore, the transmission suffers from extra waiting times for acknowledgment at the end of each short batch. Thus, as illustrated in Fig. 4, the time spent waiting for acknowledgment is larger than the time to deliver the coded packet.

Beginning from a certain SNR value, ANCEF and NC have similar performance, because in ANCEF the number of packets in each batch has been increased and converges to the exact dof value of the non-adaptive NC scheme. The behavior of ANC and ANCM is instead opposite; the goal of both is to achieve a target reliability by selecting a given number of packets per batch. It is possible to note that there is a small delay gain in ANCM over ANC, since it works by also selecting the best modulation order.

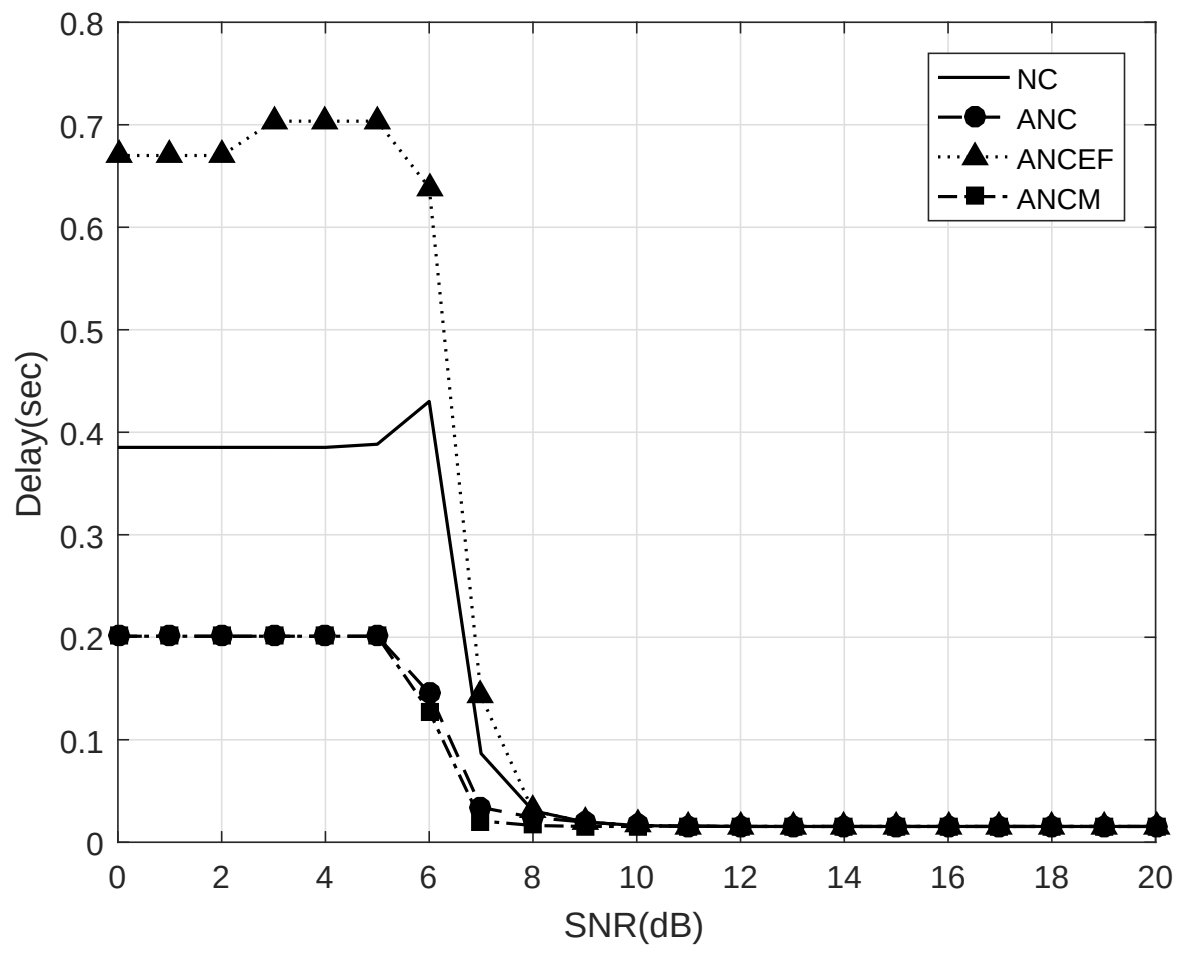

Fig. 4. Transmission Delay in a LEO Satellite scenario for variable SNR values.

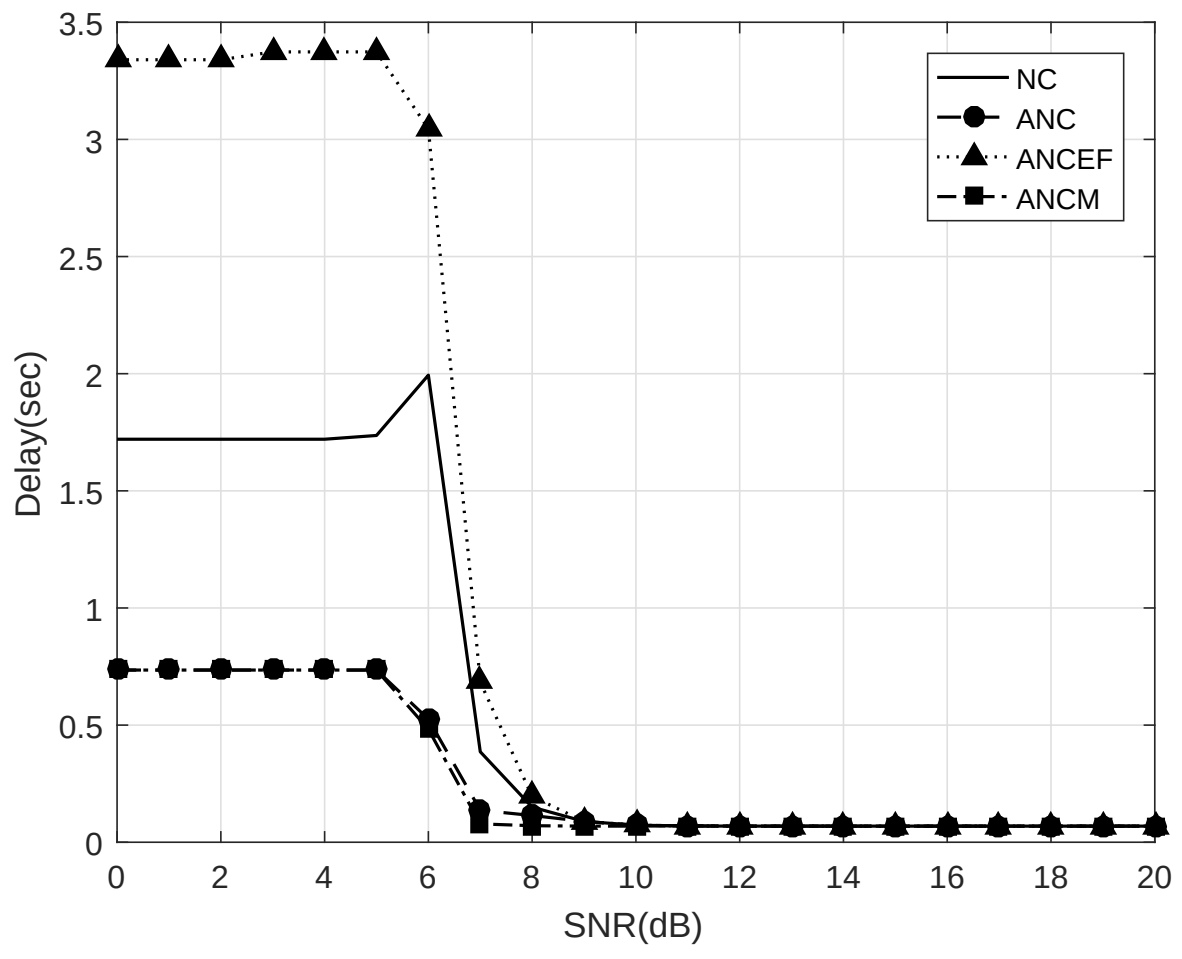

Fig. 6. Transmission Delay in a MEO Satellite scenario for variable SNR values.

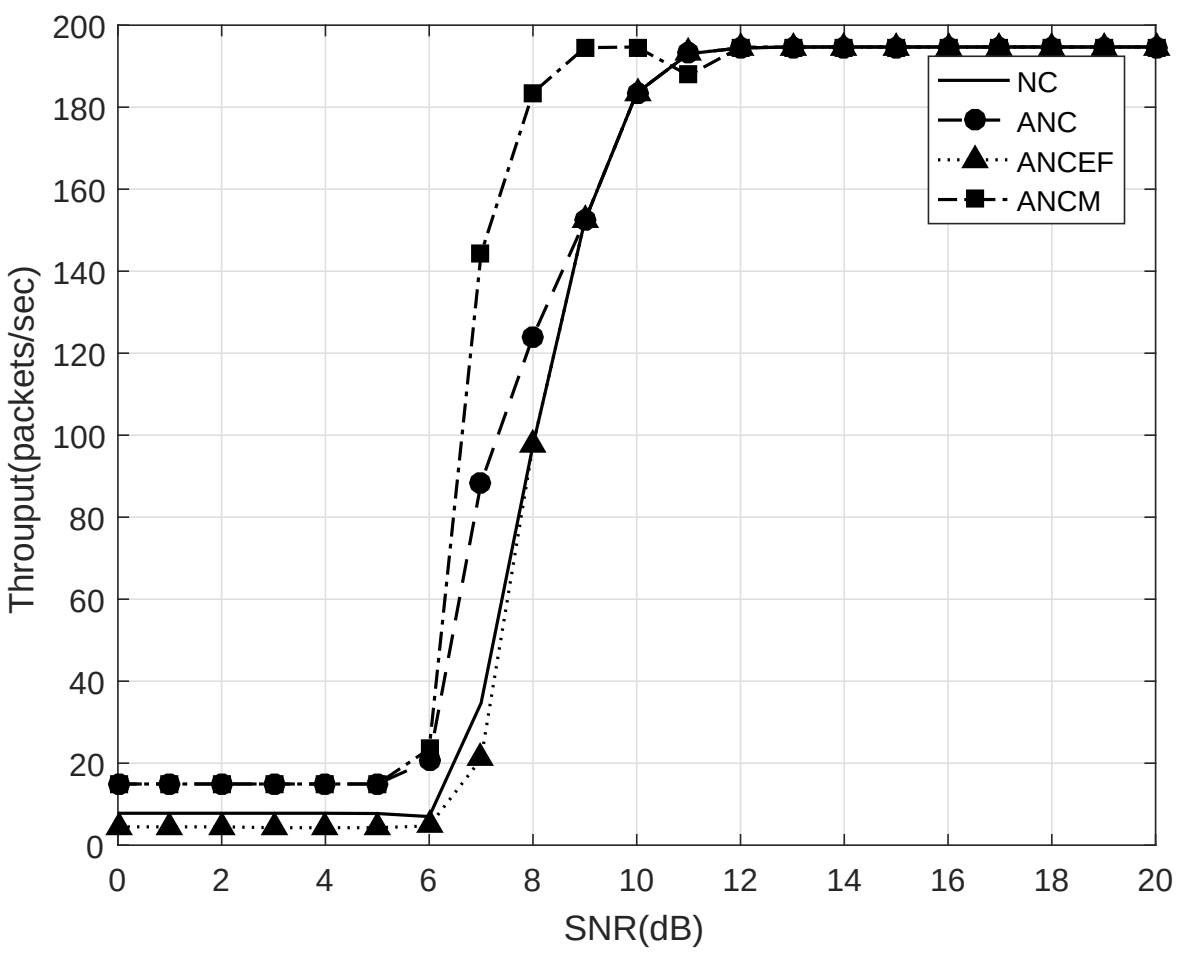

Fig. 5. Throughput in a LEO Satellite scenario for variable SNR values.

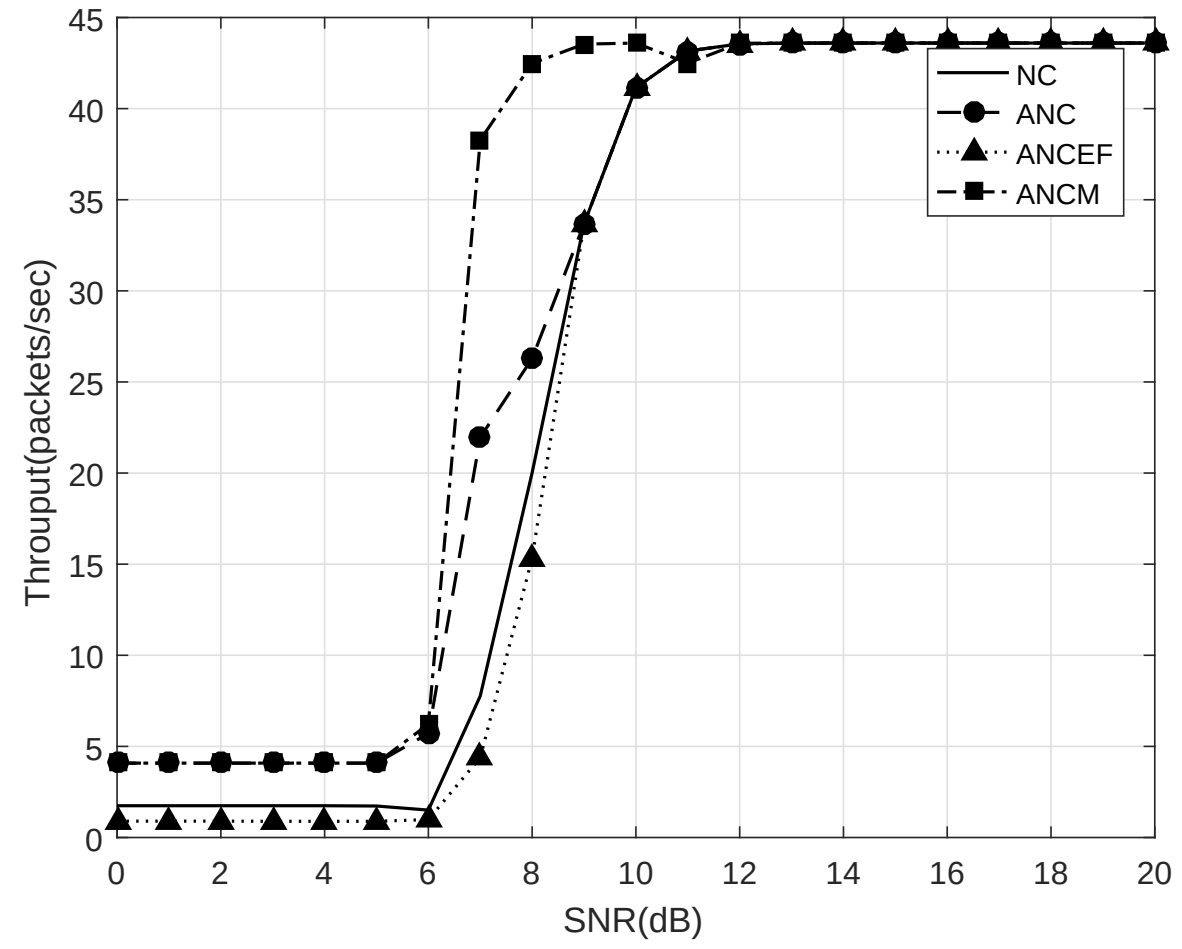

Fig. 7. Throughput in a MEO Satellite scenario for variable SNR values.

In Fig. 5 the throughput behavior is reported for both benchmark schemes, i.e., NC and ANC, and the proposed ones, i.e., ANCEF and ANCM. For low and intermediate SNR values, both ANC and ANCM have some remarkable achievements in terms of higher throughput. While at the same range of SNR, ANCEF has a very small throughput due to the reduced number of transmitted packets. However, both ANCM and ANC have gain in delay and throughput for a certain range of SNR. Anyway, it has to be stressed that the main aim of the ANCEF was to avoid energy wasting in bad channel conditions, and, hence, limit the number of sent packets.

The remaining Figs. 6-7 and Figs. 8-9 report the performance in terms of delay and throughput for MEO and GEO scenarios, respectively. Those figures prove to be similar in their trend to the ones for the LEO scenario. In particular, it is possible to see that if on one hand the behavior of the different techniques remain quite similar within a certain scenario, on the other hand the increased RTT incurs higher delay figures.

In particular, it is possible to see that the increased RTT incurs relatively higher delay for the ANCEF scheme, since it is more affected by the RTT than other schemes. Additionally, it is possible to see that the ANCM scheme, by utilizing different modulations schemes, can gain in terms of throughput in comparison to the other schemes.

## V. Conclusions

This paper addresses network coding over time varying channels. We proposed two novel adaptive physical layer aware schemes for packet transmission over time variant channels. Those schemes compensate for the lost degrees of freedom by tracking the packet erasures over time. The novelty of such schemes is that adaptation is dominated by the channel quality, therefore, at SNR values, high enough for reliable transmission, the schemes can be switched off to allow for a reduction in the processing power at the transmitter side.

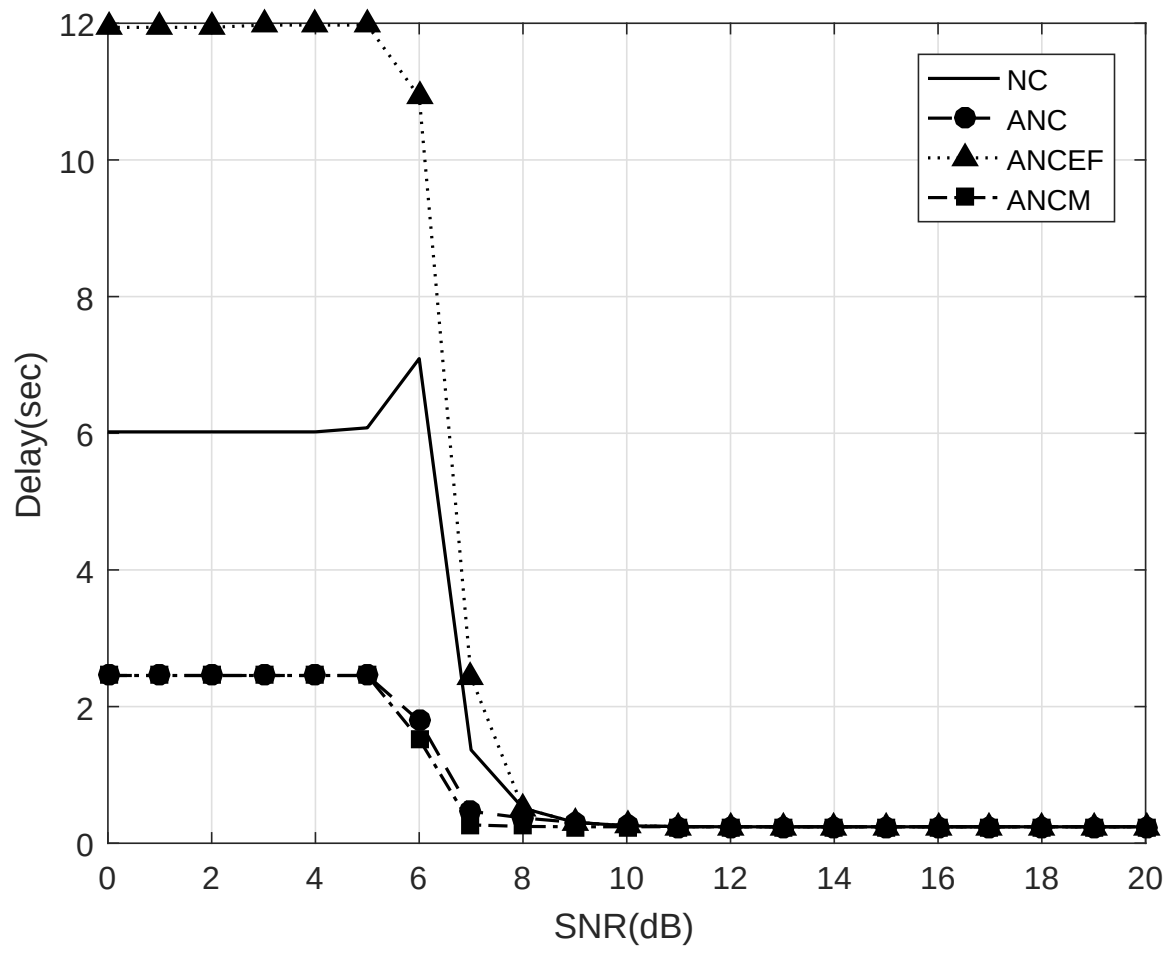

Fig. 8. Transmission Delay in a GEO Satellite scenario for variable SNR values.

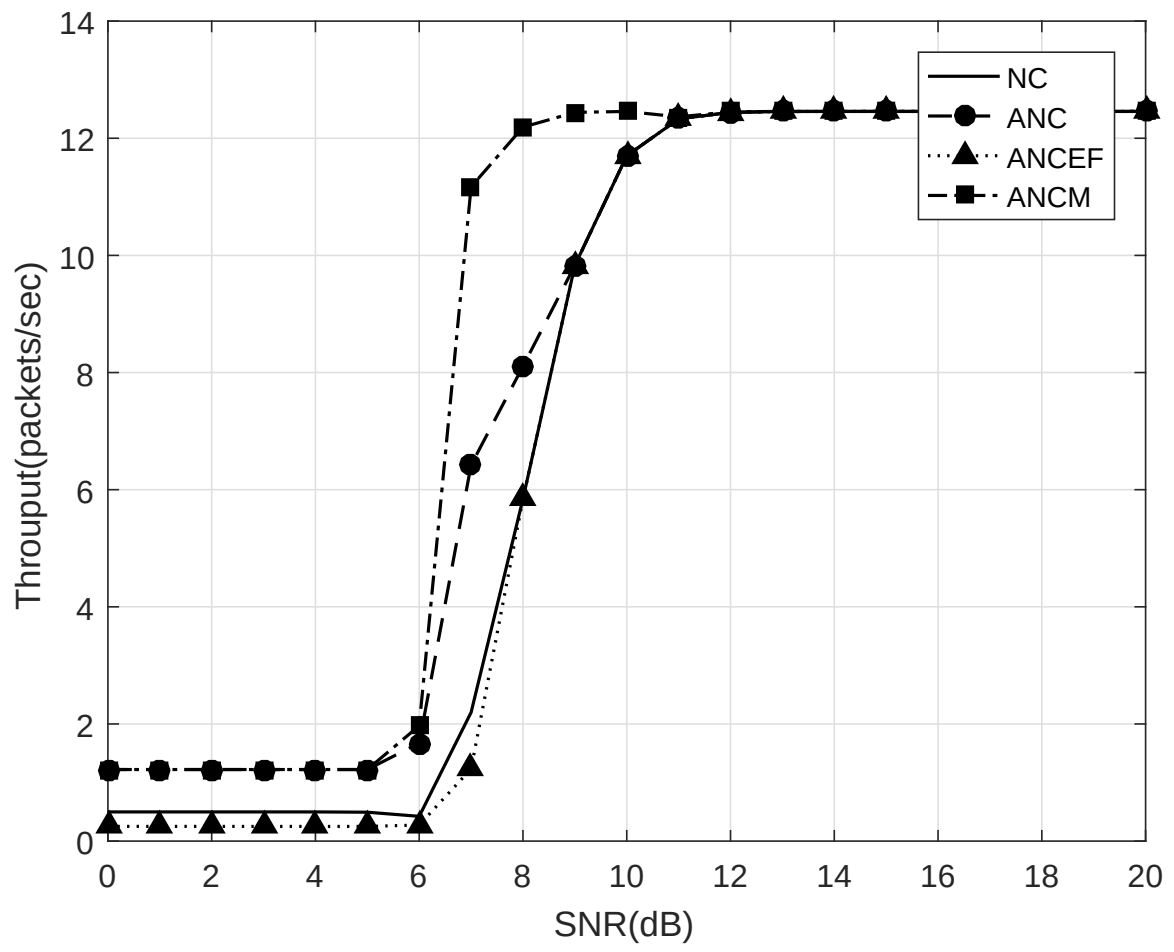

Fig. 9. Throughput in a GEO Satellite scenario for variable SNR values.

Future research will consider complexity analysis of the proposed schemes, energy analysis, considering the erasure of the acknowledgment packet and its implication on the overall modeling process and the performance of the proposed schemes. Additionally, defining transmission strategies that allows for further gains in delay-throughput or energy efficiency and finding an optimal coded packets transmission strategies to allow for such conflicting objectives is of particular interest.